# Full characterization of electronic transport properties in working polymer light-emitting diodes via impedance spectroscopy

Makoto Takada,[1] Takashi Nagase,[1,2] Takashi Kobayashi,[1,2] and Hiroyoshi Naito[1,2,a)]

[1]Department of Physics and Electronics, Osaka Prefecture University, Sakai 599-8531, Japan

[2]The Research Institute for Molecular Electronic Devices, Osaka Prefecture University, Sakai 599-8531, Japan

The electron and hole drift mobilities of organic semiconductor layers, localized tail state distributions, and bimolecular recombination constants in working polymer light-emitting diodes (PLEDs) are determined simultaneously using impedance spectroscopy (IS). The organic light-emitting layers of these PLEDs are composed of poly(9,9-dioctylfluorene-alt-benzothiadiazole) (F8BT). Electron and hole transit time effects are observed in the capacitance-frequency characteristics of the PLEDs and their drift mobilities are determined over wide temperature and electric field ranges. The drift mobilities exhibit thermally activated behavior and the localized tail state distributions from the conduction band and valence band mobility edges are then determined from analysis of the electric field dependences of the activation energies. The bimolecular recombination constants are determined from the inductive response of the impedance-frequency characteristics. The IS technique is also applicable to degradation analysis of the PLEDs; changes in the mobility balance, the localized tail state distributions, and the bimolecular recombination constant caused by aging are all shown.

[a]Corresponding author. Tel.: +81 72 254 9266; fax: +81 72 254 9266. *Electronic mail:* naito@pe.osakafu-u.ac.jp.

## I. INTRODUCTION

Organic semiconductor optoelectronic devices such as polymer light-emitting diodes (PLEDs) have been attracting considerable attention for use in optoelectronic applications because of their flexibility, light weight, and printability. To aid in the design and performance improvement of PLEDs and in understanding of the device physics, it is important to characterize the transport properties of organic semiconductors, including the electron and hole drift mobilities, the localized tail state distributions, and the recombination constants. In PLEDs, carrier transport materials with high drift mobilities and light-emitting materials with balanced electron and hole mobilities are required to reduce their driving voltages [1] and enhance their external quantum efficiencies [2, 3].

The drift mobilities in these devices are governed by the localized tail state distributions because the carrier transport in organic semiconductors is caused by incoherent carrier hopping via randomly distributed localized states. Therefore, information about the localized tail state distributions in organic semiconductors is an important aid in understanding of the optical and electronic properties of organic semiconducting materials.

The bimolecular recombination constant, which governs the recombination rate and the recombination zone, is another important physical quantity and is closely related to the external quantum efficiency of PLEDs. The bimolecular recombination process in low-mobility semiconductors (where organic semiconductors are low-mobility semiconductors) is known as Langevin recombination [4]. The Langevin recombination constant $\gamma_L$ is given by

$$\gamma_L = \frac{q}{\varepsilon}\left(\mu_n + \mu_p\right), \tag{1}$$

where $q$ is the elementary charge, $\varepsilon$ is the dielectric constant, and $\mu_n$ ($\mu_p$) is the electron mobility (hole mobility). It was reported recently that the bimolecular recombination constants of the organic semiconducting materials used in PLEDs are much lower than $\gamma_L$ [5].

The most widely used methods to measure the drift mobilities of organic semiconductors are the time-of-flight (TOF) transient photocurrent technique [6, 7] and the dark injection transient space-charge-limited current (DI SCLC) technique [8, 9]. The localized state distributions are measured using thermally stimulated current (TSC) [10], deep level transient spectroscopy (DLTS) [11], and transient photocurrent (TPC) [12] techniques. Several transient electro-optical techniques have been proposed to determine the recombination constants in working organic photovoltaics (OPVs) to date, including the transient photovoltage method [13], and photo-induced current and charge extraction by the linearly increasing voltage (photo-CELIV) method [14,15]. The photo-CELIV method can also be used to determine the drift mobilities of organic semiconductors. However, these transport properties, i.e., the electron and hole drift mobilities, the localized tail state distributions, and the bimolecular recombination constants, cannot be determined simultaneously in working PLEDs using the experimental techniques listed above.

We have previously reported that the drift mobilities, the localized tail state distributions, and the deep trapping lifetimes in organic thin-film single-injection devices can be characterized simultaneously using impedance spectroscopy (IS) [16-19]. The IS technique has proven to be a powerful tool for study of the carrier transport properties of organic devices [16-25]. The major advantages of the IS technique include its fully automatic measurement capability and its applicability to thin-film devices with thicknesses of ~100 nm, which is the typical active layer thickness in working PLEDs.

The IS technique can be applied to working double-injection devices such as PLEDs and OPVs. Numerical calculations of the impedance spectra of double-injection devices have shown that the hole and electron mobilities in these devices can be determined simultaneously from the impedance spectra when the mobilities of the two carriers differ by a factor of at least 20 [26]. Simultaneous measurements of the electron and hole mobilities have been demonstrated in both poly(p-phenylene vinylene) (PPV) polymer LEDs and tris(8-hydroxyquinolinato) aluminum (Alq$_3$) double-injection diodes [27, 28].

In this work, we show that the electron and hole mobilities, the localized state distributions from the conduction band and valence band mobility edges, and the bimolecular recombination constants of organic light-emitting semiconductors can be determined simultaneously from the impedance spectra of working PLEDs. The electron and hole mobilities of an organic light-emitting layer are measured over wide temperature and electric field ranges. The localized tail state distributions from the conduction band and valence band mobility edges are mapped out from the temperature and electric field dependences of the electron and hole mobilities, respectively. The bimolecular recombination constants are determined from the inductive responses shown in the impedance spectra. Measurements of transport properties in this manner are particularly useful in diagnosis of the degradation of organic devices. We demonstrate the applicability of IS measurements to analysis of the degradation caused by aging of working PLEDs.

## II. EXPERIMENTAL

We determined the transport properties of a green light-emitting polymer emission layer, poly(9,9-dioctylfluorene-alt-benzothiadiazole) (F8BT), in an F8BT-based inverted PLED using the IS technique. Holes are injected from a transparent metal oxide (TCO) substrate in conventional PLEDs, while the electrons are injected from the TCO substrate in inverted PLEDs. The inverted PLED structure is thus completely reversed when compared with that of a conventional PLED with respect to the TCO substrate. Because chemically reactive cathode materials are not used in inverted PLEDs, these devices are air-stable and are suitable for use in flexible device applications. F8BT is a prototypical light-emitting polymer for inverted PLEDs [25, 29, 30].

The PLED device configuration used here was AZO/PEI/F8BT/MoO$_3$/Al, where AZO is Al-doped ZnO and PEI is poly(ethyleneimine). A patterned AZO glass (0071, Geomatec) cathode was cleaned using acetone, 2-propanol and an ultraviolet (UV)-ozone treatment. Subsequently, a thin PEI layer, acting as an electron injection layer, was spun onto the AZO glass surface from an ethanol solution (0.1 wt.%, 2000 rpm, 30 s). The substrate was then annealed in the ambient atmosphere (5 min, 150°C). A 270-nm-thick F8BT layer was spun onto the PEI layer from a chlorobenzene solution to act as an emissive layer (1.2 wt.%, 800 rpm, 60 s). After F8BT emissive layer deposition, the substrates were dried at 80°C for 30 min. 10-nm-thick MoO$_3$ and 50-nm-thick Al layers were then successively thermally evaporated onto the F8BT emissive layer in a vacuum chamber at a base pressure of $10^{-3}$ Pa. Finally, the PLEDs were encapsulated using epoxy. The active area of each fabricated device was 4 mm$^2$.

The current density-voltage (J-V) characteristics of the PLEDs were recorded using a source measurement unit (Model 2411, Keithley). The luminance was measured using a luminance meter (CS-200, Konica Minolta). The impedance spectra were measured using a Solartron ModuLab XM over the range from $10^0$ to $10^6$ Hz while the devices were held in a probe station (TTP-4, Desert Cryogenics) at temperatures ranging from 100 to 300 K.

### III. RESULTS AND DISCUSSION

The J-V and luminescence-voltage (L-V) characteristics of the PLEDs are shown in Fig. 1. The luminescence at 5.0 V is 6,600 cd m$^{-2}$ (at 96 mA cm$^{-2}$). The current efficiency versus current density characteristics of the PLEDs obtained from Fig. 1 are shown in Fig. 2. The maximum current efficiency is 7.2 cd A$^{-1}$ (at 20 mA cm$^{-2}$), which is nearly comparable to the efficiency values reported in the literature [25, 31-33].

The method used to measure the carrier drift mobility in either an electron-only device (EOD) or a hole-only device (HOD) is known as the $-\Delta B$ method [34], where $-\Delta B$ is the negative differential susceptance. Note that only electrons (holes) are injected in EODs (HODs). In this method, $-\Delta B$ is given by

$$-\Delta B = -\omega \big[ C(\omega) - C_{geo} \big],\tag{2}$$

where $\omega$ is the angular frequency (= $2\pi f$), $C$ is the capacitance and $C_{geo}$ is the geometrical capacitance. The frequency $f_{max}$ at the maximum in the $-\Delta B$ spectrum is related to the carrier transit time $t_t$ by

$$t_t \approx 0.72 f_{max}^{-1}.\tag{3}$$

The carrier drift mobility $\mu_d$ can then be determined from $t_t$ using Eq. (3) [35].

$$\mu_d = \frac{4}{3} \frac{d^2}{t_t (V_{dc} - V_{bi})} \approx \frac{4}{3} \times \frac{f_{max}}{0.72} \frac{d^2}{(V_{dc} - V_{bi})}, \tag{4}$$

where $d$ is the semiconductor layer thickness, $V_{dc}$ is the applied dc voltage, and $V_{bi}$ is the built-in potential.

It has been shown numerically that the electron and hole drift mobilities in double-injection devices can be determined simultaneously [26]. Simultaneous determination of the electron and hole mobilities is possible under the following three conditions [26]: 1) the injection barrier heights for both the anode and the cathode in double-injection devices must be low enough (<0.1 eV) to allow simultaneous injection of both electrons and holes at comparable rates; 2) the recombination rate between the electrons and holes must be in a range from $10^{-1}$ to $10^{-3}$ smaller than $\gamma_L$, and no transit time effects should be observed in the case of strong recombination ($>\gamma_L$); and 3) the trap concentrations must be sufficiently low to ensure that their contributions to the impedance spectra do not mask the susceptance peak that corresponds to the carrier with the lowest mobility.

The $C$-$f$ characteristics of the PLEDs measured under various applied voltages at 180 K are shown in Fig. 3(a). Two stepwise increases in the $C$-$f$ characteristics at $V_{dc} = 3.0$ V can be observed in the frequency ranges from 5 to 10 Hz and from 100 to 1000 Hz; these increases are caused by transit time effects [34]. The frequency regions in which the two transit time effects are observed shift toward higher frequencies as the applied voltage increases. The $C$-$f$ characteristics of the PLEDs at various temperatures at $V_{dc} = 5.0$ V are shown in Fig. 3(b). A shift in the transit time effects with increasing temperature, similar to that observed with increasing applied voltage in Fig. 3(a), is again observed.

The $-\Delta B$-$f$ characteristics of the PLEDs at the various applied voltages and temperatures that were obtained from Fig. 3(a) and 3(b) are shown in Fig. 4(a) and 4(b), respectively. The two peaks observed in the $-\Delta B$-$f$ characteristics are caused by the transit times of the electrons and the holes. We determined the electron and hole drift mobilities ($\mu_n$ and $\mu_p$, respectively) of F8BT in the PLEDs from Fig. 4 using Eqs. (3) and (4). The temperature dependence of $\mu_n$ at various applied voltages is shown in Fig. 5(a) and the corresponding dependence of $\mu_p$ is shown in Fig. 5(b). The values of $\mu_n$ and $\mu_p$ for F8BT in Fig. 5 can be identified using IS measurements in EODs and HODs based on F8BT, respectively [16, 18, 19]. The $\mu_n$ and $\mu_p$ values of F8BT at 300 K that were estimated from Fig. 5 are consistent with the corresponding mobilities of F8BT reported in the literature [36-38].

The temperature dependences of both $\mu_n$ and $\mu_p$ of F8BT in Fig. 5 show thermally activated behavior, which can be described using

$$\mu_n, \mu_p \propto \exp\left(-\frac{E_a}{kT}\right),\tag{5}$$

where $E_a$ is the activation energy, $k$ is the Boltzmann constant, and $T$ is the temperature. The $E_a$ values of $\mu_n$ and $\mu_p$ for F8BT that can be obtained from Fig. 5 are dependent on the applied electric field and vary in the ranges from $0.067{-}0.31$ eV ($E_a$ of $\mu_n$) and $0.081{-}0.47$ eV ($E_a$ of $\mu_p$).

The temperature dependences of the drift mobilities can be used to derive localized tail state distributions. The thermally activated carrier transport behavior in disordered semiconductors can be explained based on the presence of continuously decaying localized states from a mobility edge [39, 40]. It was previously shown that the localized tail state distributions $N_t$ can be obtained from the values of $E_a$ at various electric fields [41]. To do this, $E$ is used as the depth of the localized states below the conduction band edge $E_c$, and it is assumed that the trapping time of the localized states below $E_{th}$, denoted by $\tau_C$, is equal to the transit time of the microscopic mobility $t_0$ ($t_0 = d / \mu_m F$):

$$\tau_C \approx t_0 = \left[c_n \int_{E_{th}}^{\infty} N_t(E)\,dE\right]^{-1},\tag{6}$$

where $\mu_m$ is the microscopic mobility of the electrons, $F$ is the electric field, $c_n$ is the electron capture coefficient, and $E_{th}$ is the thermalization energy, which is given by

$$E_{th} = kT\ln(\nu t),\tag{7}$$

where $\nu$ is the attempt-to-escape frequency of the localized states and $t$ is the time after pulsed photoexcitation in the TOF experiments. $E_{th}$ is the demarcation energy, which separates the localized states above $E_{th}$, where the electrons are thermalized as a result of multiple thermal excitations, from the deeper states in which the electrons are unlikely to be thermally released in time $t$. The observed $E_a$ can be determined using the depth of $E_{th}$ (where $E_a = E_{th}(t_t)$), and an increase in $t_t$ should therefore cause an increase in $E_a$ [39].

Combination of $t_0 = d / \mu_m F$ with Eq. (6) yields

$$\int_{E_{th}}^{\infty} N_t(E)\,dE = (\mu_m / c_n d)F.\tag{8}$$

Equation (8) can be expressed as

$$N_t(E_{th}) = -(\mu_m / c_n d)dF / dE_{th},\tag{9}$$

where $E_{th}$ corresponds to $E_a$. Equation (9) was derived for the electron transport case, but the same equation is applicable to analysis of the hole transport; in addition, the equation was derived for the case of time-domain experiments (TOF experiments), but it is also applicable to analysis of data obtained from frequency-domain experiments (IS measurements) [42, 43] (see Supplemental Information).

The localized tail state distributions from the conduction band and valence band mobility edges of the F8BT layer in the PLEDs, which were determined using Eq. (9) from the electric field dependences of $E_a$ obtained from Fig. 5, are shown in Fig. 6. The localized tail state distributions are described well by the following Gaussian distributions:

$$N(E) = \frac{N_0}{\sqrt{2\pi}\sigma} \exp\left[-\frac{E^2}{2\sigma^2}\right],$$ (10)

where $N_0$ is the total density of localized states and $\sigma$ is the width of the density of states. The widths of the Gaussian tail states obtained from the conduction band and valence band edges of F8BT, denoted by $\sigma_n$ and $\sigma_p$, are 0.14 eV and 0.17 eV, respectively.

The method used to determine the bimolecular recombination constants $\beta$ is based on an analytical solution for the small-signal ac impedance of double-injection space-charge-limited diodes [44]. In double-injection devices such as OLEDs, the electrons and holes are injected into the light-emitting layers. In Ref. 24, we derived an analytical expression for the small-signal impedances of double-injection devices in the presence of bimolecular recombination to explain the inductive response (i.e., the negative capacitance) observed in the capacitance-frequency characteristics of the OLEDs [22]. The small-signal ac impedance $Z$ is

$$Z = 3R_0 \sum_{k=0}^{\infty} \frac{1}{k+3} \frac{\left(-\frac{3}{2}j\omega C_{geo}R_0\right)^k \left(2 + \frac{j\omega}{\beta n_0}\right)^{k+1}}{\left(3 + \frac{j\omega}{\beta n_0}\right)\left(3 + \frac{j\omega}{\beta n_0} + 1\right)\cdots\left(3 + \frac{j\omega}{\beta n_0} + k\right)},$$ (11)

where $R_0$ is the resistance of the semiconductor thin film (where $R_0 = V_0/I_0$), $V_0$ is the dc voltage component, $I_0$ is the dc component of the current ($I_0 = J_0 S$), $S$ is the active area, and $n_0$ is the steady-state carrier density. At sufficiently low frequencies (i.e., $3/2\, j\omega C_{geo}R_0 \ll 1$), the first term in Eq. (11) is dominant, and the diode impedance thus becomes

$$Z = R_0 \frac{2 + j\omega/\beta n_0}{3 + j\omega/\beta n_0}.$$ (12)

The imaginary part of Eq. (12) is

$$\text{Im}[Z] = R_0 \frac{\omega/\beta n_0}{9 + \omega^2/\beta^2 n_0^2} . \tag{13}$$

By differentiating $\text{Im}[Z]$ with respect to $\omega$, we obtain $\omega_0 = 3\beta n_0$ at $d\,\text{Im}[Z]/d\omega = 0$, and we can thus obtain a value of $\beta$ from $f_0$ at the maximum of $\text{Im}[Z]$:

$$\beta = \frac{\omega_0}{3n_0} = \frac{2\pi}{3n_0} f_0 . \tag{14}$$

$f_0$ is observed when $\beta$ is smaller than $\gamma_L$, i.e., when the capacitance is negative [24]. When the capacitance is not negative, $\beta$ cannot be determined and has a value greater than or equal to $\gamma_L$. Here, we estimate the value of $n_0$ from the $J$-$V$ characteristics [44] of working double-injection devices using

$$J = q(\mu_n + \mu_p)n_0 F . \tag{15}$$

The $\text{Im}[Z]$-$f$ characteristics of the PLEDs at various applied voltages are shown in Fig. 7, in which inductive responses (negative capacitances) are observed. The electric field dependences of the $\beta$ values of the light-emitting layer are determined from the $f_0$ observed in Fig. 7 using the values of $n_0$ [in Eq. (15)] and these dependences are shown in Fig. 8. The $\gamma_L$ values determined from $\mu_n$ and $\mu_p$ in Fig. 5 are also shown in Fig. 8.

The Langevin recombination constant $\gamma_{Ld}$ in disordered semiconductors with distributed tail states from their mobility edges is also expressed as per Eq. (1), in which $\mu_n$ is the electron drift mobility and $\mu_p$ is the hole drift mobility [45].

The value of $\beta$ at 298 K and $8.0 \times 10^4$ V cm$^{-1}$ is $4.3 \times 10^{-12}$ cm$^3$s$^{-1}$, which is approximately $10^2$ times smaller than $\gamma_L$. Plots of $\beta$ and $\gamma_L$ versus temperature at $1.0 \times 10^5$ V cm$^{-1}$ are shown in Fig. 9. The temperature dependences of both $\beta$ and $\gamma_L$ exhibit thermally activated behavior. The activation energy value of $\beta$ is 0.26 eV, which is the same as the corresponding value of $\gamma_L$. Plots of the reduction factor for Langevin recombination ($\beta/\gamma_L$) versus temperature in the PLEDs under various applied voltages are shown in Fig. 10. The $\beta/\gamma_L$ values of the as-prepared PLEDs are in the $10^{-3}$–$10^{-2}$ range and are temperature-independent. In OPVs, the $\beta/\gamma_L$ value is dependent on temperature and this temperature dependence can be explained as the effect of random spatial fluctuations in the potential landscapes of the disordered semiconductors [46]. The origin of the temperature-independent reduction factors is not known at present. Therefore, the bimolecular recombination process for F8BT in PLEDs can be described as a Langevin-like recombination model [45] with a temperature-independent reduction factor [47].

The organic device performance degradation caused by aging is a serious problem for commercialization of these devices. IS measurements can provide important knowledge of the changes in the transport properties of working PLEDs after aging. As an example, we present the results of degradation analyses performed in the PLEDs. The temporal variations in the driving voltage and the luminance of the PLEDs operating at 20 mA cm$^{-2}$ are shown in Fig. 11. The driving voltage increases from 4.0 to 4.8 V while the luminance decreases from 1,700 to 400 cd m$^{-2}$ after driving for 12 h. The *J-V* and *L-V* characteristics of the 12-h-aged PLEDs are shown in Fig. 1 and their current efficiency-current density characteristics are shown in Fig. 2. The current efficiency at 20 mA cm$^{-2}$ in the 12-h-aged PLEDs is 2.3 cd A$^{-1}$, which is 3.1 times lower than that in as-prepared PLEDs.

The temperature dependences of the $\mu_n$ and $\mu_p$ values of the light-emitting layer in the 12-h-aged PLEDs are shown in Fig. 12. The $E_a$ values of both the electron and hole drift mobilities are increased by aging under constant current operation conditions. The $\mu_n$ value at 300 K in the 12-h-aged PLEDs is almost the same as that in the as-prepared PLEDs (the $\mu_n$ values at 300 K and 1.6×10$^5$ V cm$^{-1}$ are 9.1×10$^{-4}$ cm$^2$V$^{-1}$s$^{-1}$ in the as-prepared PLEDs and 7.5×10$^{-4}$ cm$^2$V$^{-1}$s$^{-1}$ in the 12-h-aged PLEDs). The $\mu_p$ value at 300 K in the 12-h-aged PLEDs is lower than that in the as-prepared PLEDs (the $\mu_p$ values at 300 K and 1.6×10$^5$ V cm$^{-1}$ are 5.8×10$^{-5}$ cm$^2$V$^{-1}$s$^{-1}$ in the as-prepared PLEDs and 1.5×10$^{-5}$ cm$^2$V$^{-1}$s$^{-1}$ in the 12-h-aged PLEDs). The localized tail state distributions in the 12-h-aged PLEDs from the electric field dependence of the $E_a$ values are presented in Fig. 6. The $\sigma_n$ value of 0.19 eV and $\sigma_p$ value of 0.22 eV for F8BT in the 12-h-aged PLEDs obtained by fitting Eq. (10) to the localized tail state distributions are both higher than the corresponding values in the as-prepared PLEDs.

The electric field dependences of the $\beta$ and $\gamma_L$ values of the light-emitting layer in the 12-h-aged PLEDs are shown in Fig. 8. The $\beta$ value at 298 K in the 12-h-aged PLEDs is lower than that in the as-prepared PLEDs (the $\beta$ value at 298 K and 8.0×10$^4$ V cm$^{-1}$ in the 12-h-aged PLEDs is 8.9×10$^{-13}$ cm$^3$s$^{-1}$). Plots of $\beta$ and $\gamma_L$ versus temperature at 1.0×10$^5$ V/cm in the 12-h-aged PLEDs are shown in Fig. 9. The activation energy of $\beta$ in the 12-h-aged PLEDs is 0.34 eV, while that of $\gamma_L$ is 0.31 eV. The activation energies of $\beta$ in the 12-h-aged PLEDs are increased by aging during constant current operation and are slightly higher than the corresponding values for the activation energy of $\gamma_L$ in the 12-h-aged PLEDs.

Plots of the Langevin recombination reduction factor $\beta/\gamma_L$ versus temperature under various applied voltages in the as-prepared and the 12-h-aged PLEDs are shown in Fig. 10. The $\beta/\gamma_L$ values in the as-prepared PLEDs and the 12-h-aged PLEDs are both independent of $\sigma$ in the presence of Gaussian distributions of the tail states, as expressed using Eq. (10); however, the $\sigma_n$ and $\sigma_p$ values of F8BT in the 12-h-aged PLEDs are higher than the corresponding values in the as-prepared PLEDs. Monte Carlo simulations showed that $\beta/(\mu_n + \mu_p)$, i.e., $\beta/\gamma_L$, is independent of the widths of the Gaussian distributions of the tail states [48]. Therefore, the results shown in Fig. 10 are consistent with the simulation results presented in Ref. 48.

Plots of the current efficiency versus the ratio of the bimolecular recombination rate to the current density ($\beta np/J$) in the as-prepared and 12-h-aged PLEDs are shown in Fig. 13, where $\beta np$ can be regarded as the calculated emission intensity, and $\beta np/J$ is thus a physical quantity that corresponds to the current efficiency (meaning that $\beta np/J$ is thereby proportional to the current efficiency above $10^{21}$ $A^{-1}cm^{-1}s^{-1}$ in Fig. 13). The current efficiency for $\beta np/J$ in the 12-h-aged PLEDs is lower than that in the as-prepared PLEDs, which indicates that the nonradiative recombination rate of the electron-hole pairs increases with aging.

Aging of the PLEDs thus enhances the degree of disorder of F8BT, which is manifested in the increased widths of the localized tail state distributions shown in Fig. 6. The degradation of aged PLEDs is mainly caused by their reduced mobility balance [2, 3] and the increased nonradiative recombination of the electron-hole pairs, based on the experimental results shown in Figs. 6, 8, 9, 10, and 13.

As shown above, the $\mu_n$, $\mu_p$ and $\beta$ values of the light-emitting layer can be determined simultaneously using IS in working PLEDs. Determination of the $\mu_n$ and $\mu_p$ values of organic semiconductors in working devices also provides information about their localized tail state distributions. These transport properties are useful in studies of PLED degradation, as demonstrated above. In addition, the transport properties measured in the working devices can be used as inputs for device simulations. These device simulations will enable device performance prediction and thus allow the device structures to be optimized to obtain maximum device performance.

## IV. CONCLUSIONS

We have demonstrated simultaneous determination of the electron and hole drift mobilities and bimolecular recombination constants of working PLEDs with F8BT emissive layers by the IS technique. The values of the electron and hole drift mobilities of F8BT are consistent with previously reported values in the literature. We show that the localized tail state distributions in F8BT can be determined based on the electric field dependences of the activation energies of the carrier drift mobilities. The localized tail state distributions from the conduction band and valence band mobility edges in F8BT can be described well using Gaussian distributions. The values of the bimolecular recombination constants are $10^2$ times lower than corresponding values of the Langevin recombination constants determined using the drift mobilities. However, the activation energy for the bimolecular recombination constants is the same as that for the Langevin recombination constants, as calculated from Eq. (1), which means that the bimolecular recombination of F8BT in the PLEDs can be explained using a Langevin-like recombination model with a temperature-independent reduction factor. We also demonstrated the applicability of the IS technique to analysis of the degradation observed in F8BT PLEDs, which was caused by continuous device driving. The changes in the transport properties in the F8BT PLEDs, including the reduction in the hole drift mobility, the increased widths of the localized tail state distributions, and the aging-induced reduction in the bimolecular recombination constants can be observed clearly using IS. In addition, the transport properties determined using IS are useful as inputs for device simulations, which will aid in the design of PLED structures and in understanding of the underlying device physics.

## SUPPLEMENTARY MATERIAL

See the supplementary material for calculation of localized state energy distributions from the electric-field dependence of the activation energies of the drift mobilities determined by means of IS.

## ACKNOWLEDGMENTS

This work was partly supported by JSPS KAKENHI (Grant Number JP17H01265), by the Murata Science Foundation (Grant Number H29RS72), and by the ICOM Foundation, Japan. The authors would like to thank Sumitomo Chemical for supplying the F8BT and Nippon Shokubai for supplying the PEI. We thank David MacDonald, MSc, from Edanz Group (www.edanzediting.com/ac) for editing a draft of this manuscript.

## V. FIGURES

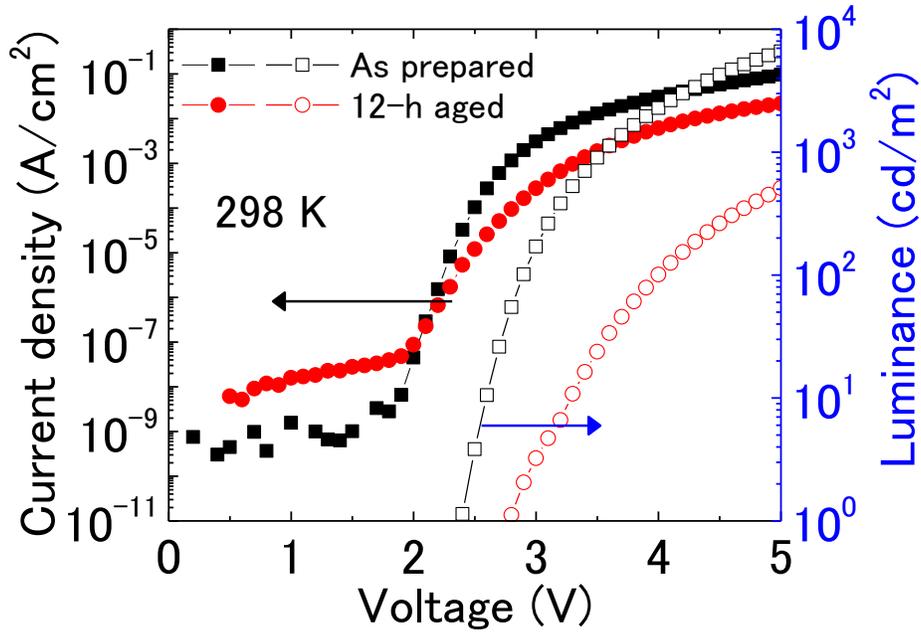

**FIG 1.** Current density-voltage (closed symbols) and luminescence-voltage (open symbols) characteristics of F8BT-based PLEDs, where the device configuration was AZO/PEI/F8BT/MoO₃/Al (squares: as-prepared PLEDs; circles: 12-h-aged PLEDs).

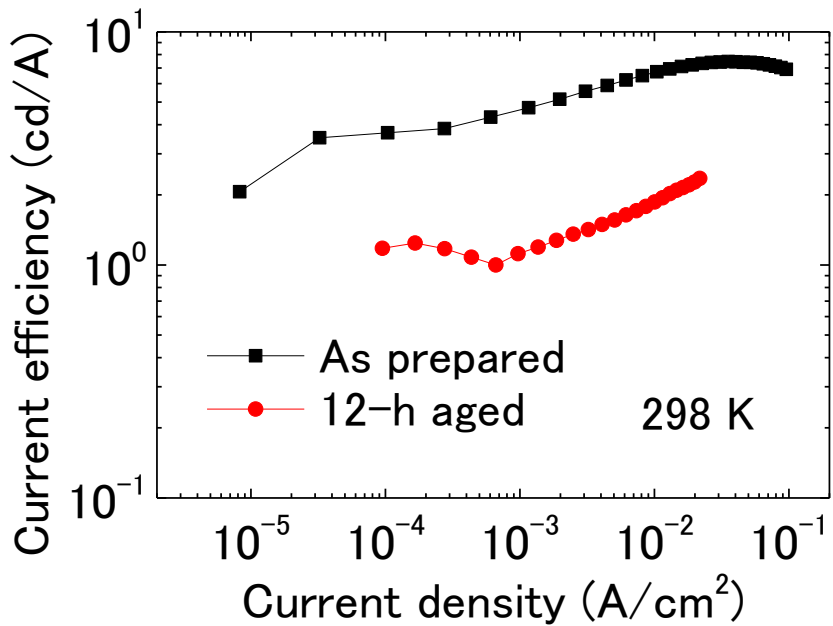

**FIG 2.** Plots of current efficiency versus current density in F8BT-based PLEDs as obtained from Fig. 1 (squares: as-prepared PLEDs; circles: 12-h-aged PLEDs).

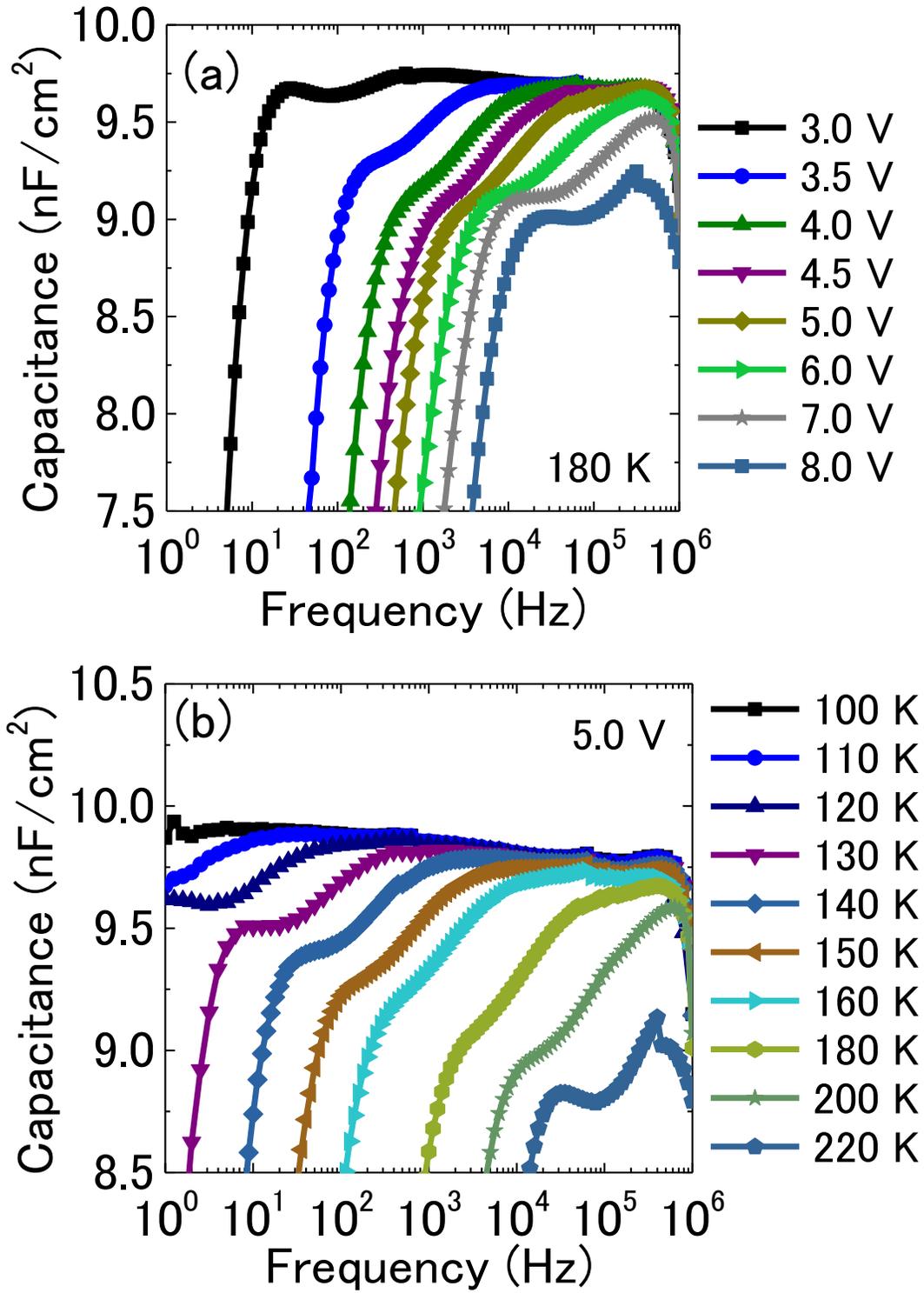

**FIG 3.** Frequency dependences of capacitance in F8BT-based PLEDs for (a) various applied voltages at 180 K and (b) various temperatures at $V_{dc} = 5.0$ V. The two transit time effects of electrons and holes are observed in the $C$-$f$ characteristics.

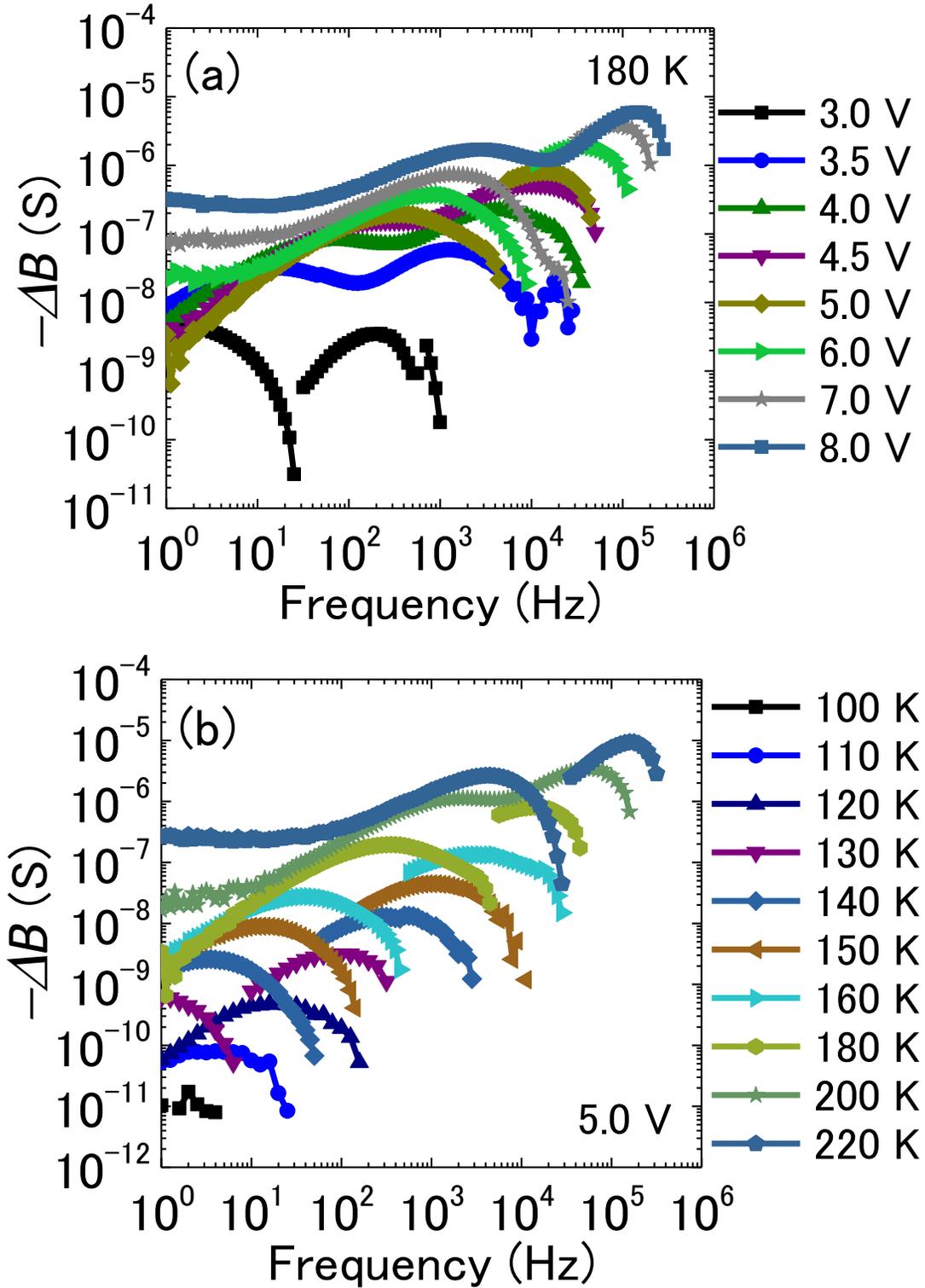

**FIG 4.** Frequency dependences of $-\Delta B$ in F8BT-based PLEDs for (a) various applied voltages at 180 K and (b) various temperatures at $V_{dc}$ = 5.0 V. Two peaks in $-\Delta B$ that are due to the transit times of the electrons and holes are observed in the $-\Delta B$-$f$ characteristics.

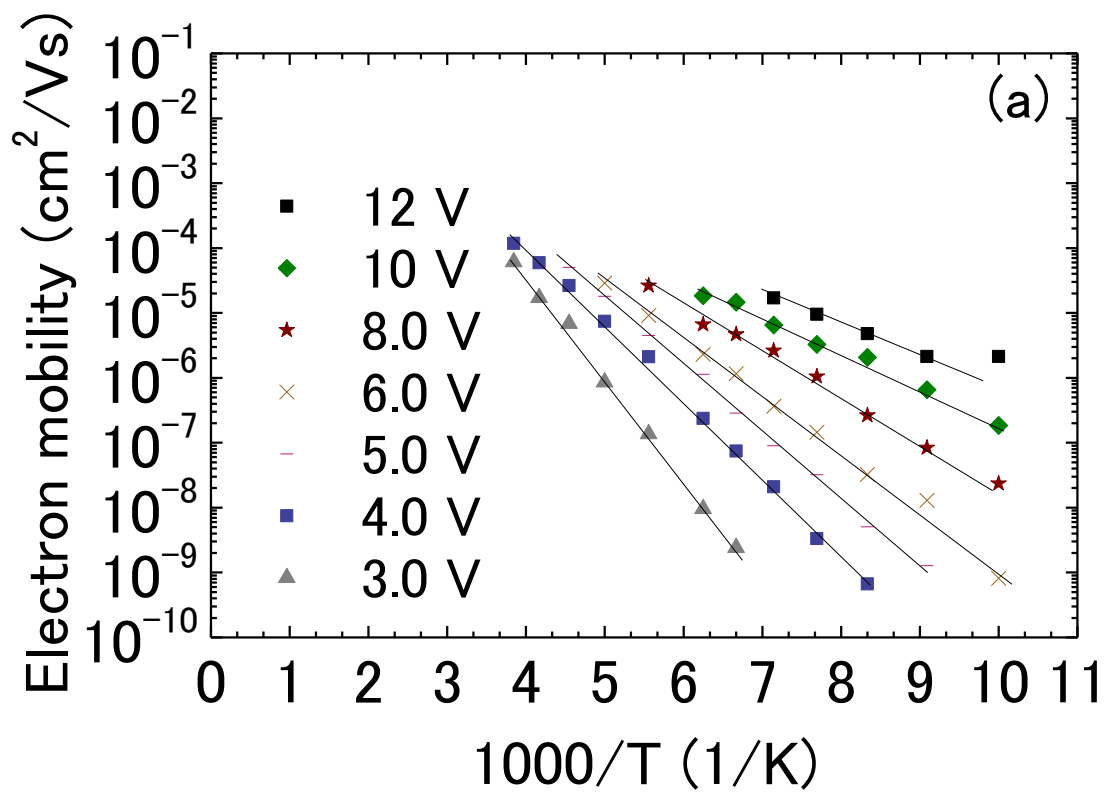

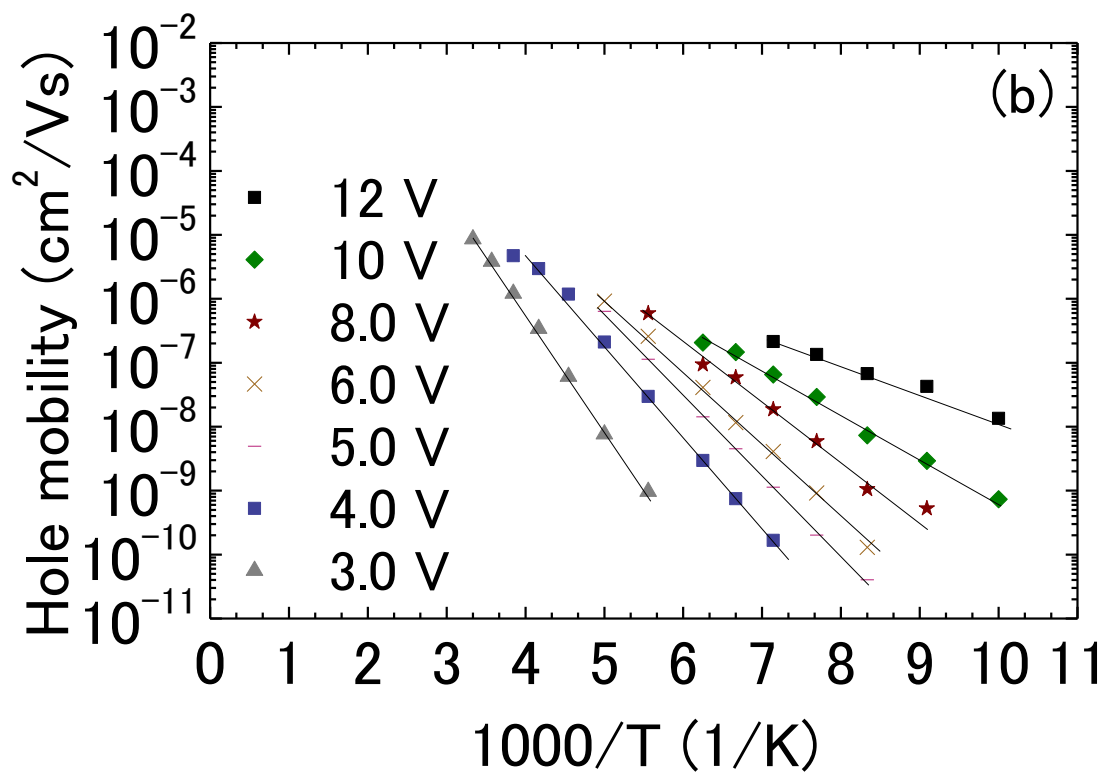

**FIG 5.** Plots of (a) electron drift mobility ($\mu_n$) and (b) hole drift mobility ($\mu_p$) versus temperature under various applied voltages in F8BT-based PLEDs.

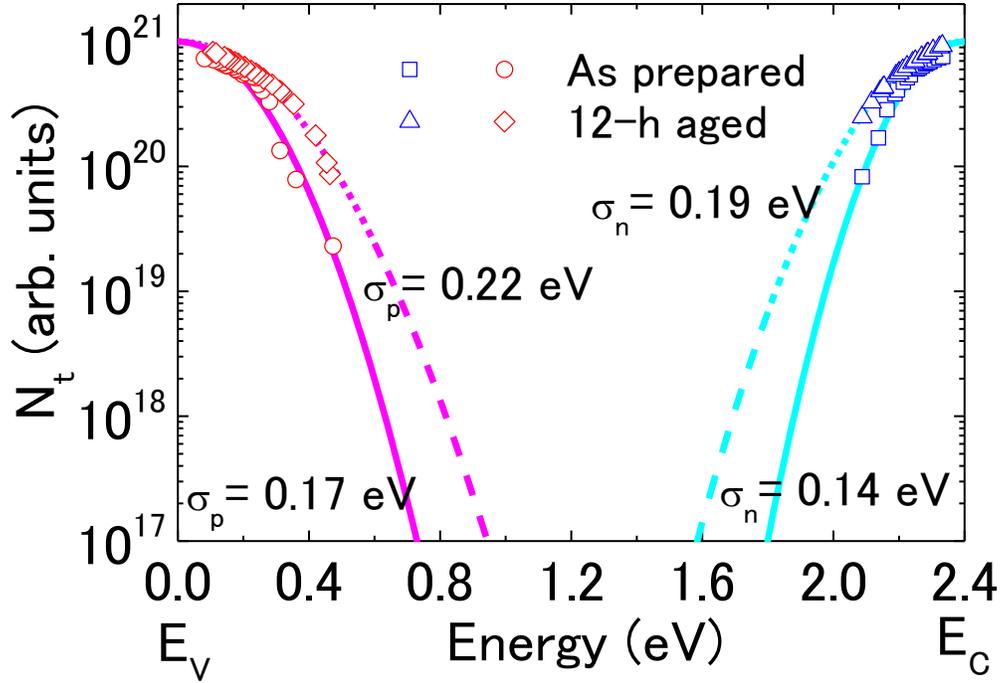

**FIG 6.** Localized tail state distributions from the conduction band and valence band mobility edges of the F8BT layers in as-prepared (conduction band: squares; valence band: circles) and 12-h-aged (conduction band: triangles; valence band: diamonds) PLEDs, as determined from the electric field dependences of $E_a$ obtained using Eq. (9) from the $\mu_n$ and $\mu_p$ values of F8BT in Figs. 5 and 12. The localized state distributions can be described well using Gaussian distributions (solid lines: as-prepared PLEDs; dashed lines: 12-h-aged PLEDs) and the widths of the Gaussian tail states obtained from the conduction band edge ($\sigma_n$) and the valence band edge ($\sigma_p$) in the as-prepared PLEDs are $\sigma_n = 0.14$ eV and $\sigma_p = 0.17$ eV. Both $\sigma_n$ and $\sigma_p$ are increased by aging under constant current driving; in the 12-h-aged PLEDs, $\sigma_n$ is 0.19 eV and $\sigma_p$ is 0.22 eV.

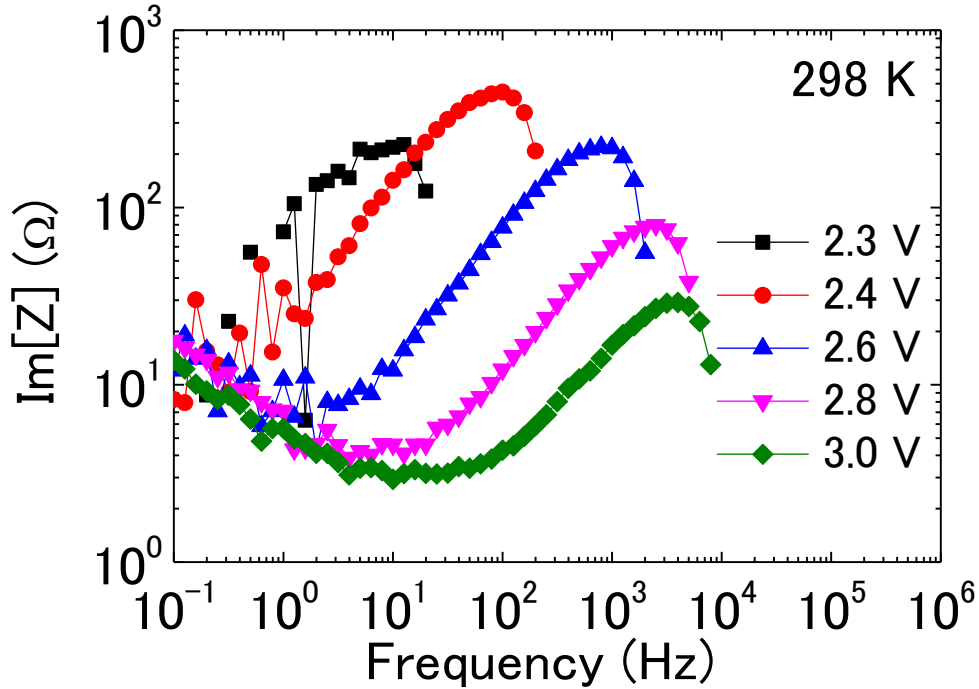

**FIG 7.** Frequency dependences of the imaginary part of the impedance (Im[$Z$]) in F8BT-based PLEDs under various applied voltages at 298 K. The maximum of Im[$Z$] is observed in the Im[$Z$]-$f$ characteristics.

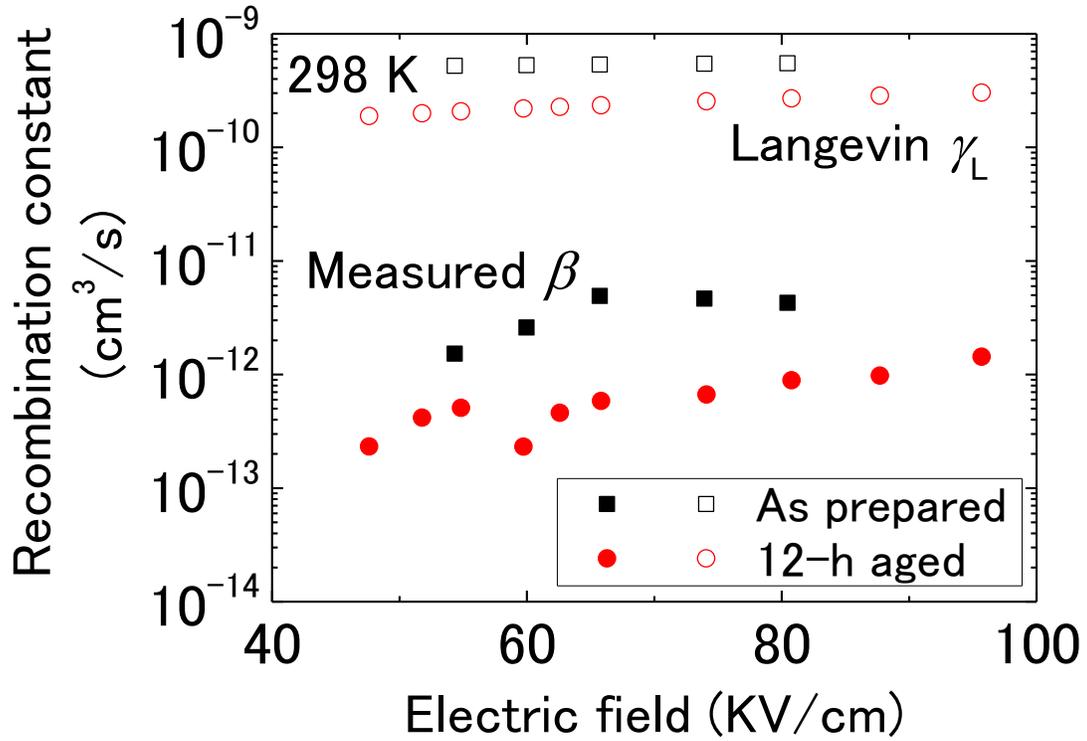

**FIG 8.** Electric field dependences of the recombination constant $\beta$ of F8BT determined from the frequency of the maximum in Im[$Z$] shown in Fig. 7 (closed symbols) and the Langevin recombination constant $\gamma_L$ (open symbols) (squares: as-prepared PLEDs; circles: 12-h-aged PLEDs).

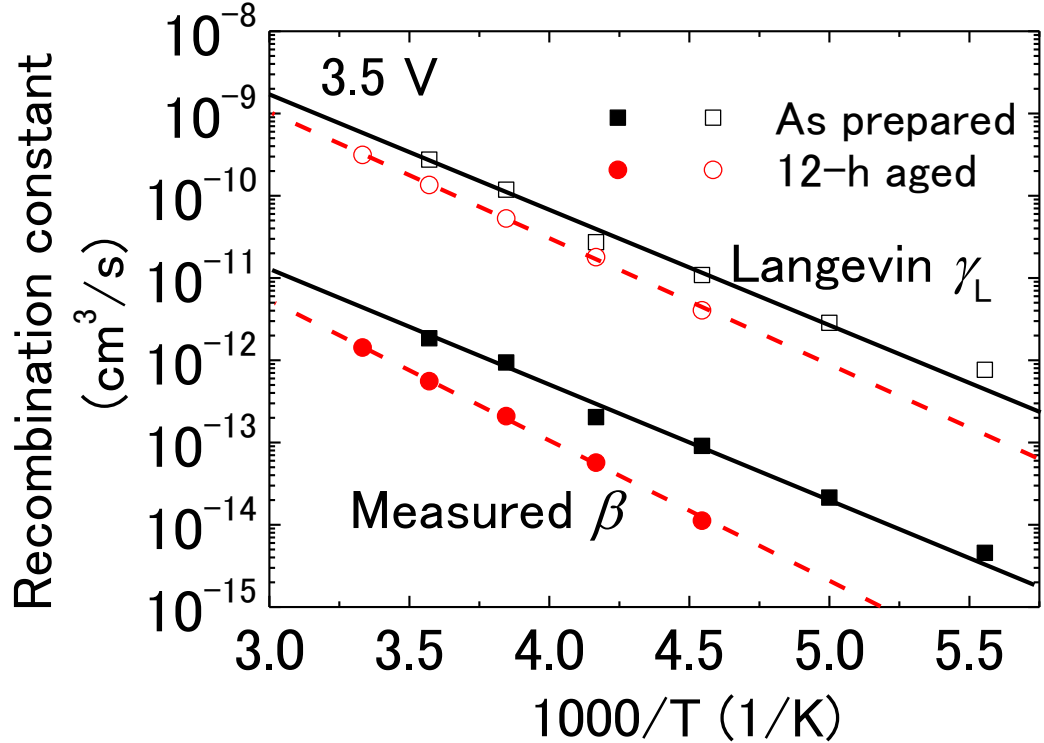

**FIG 9.** Plots of the bimolecular recombination constant $\beta$ (closed symbols) and the Langevin recombination constant $\gamma_L$ (open symbols) versus temperature at $1.0 \times 10^5$ V/cm in the as-prepared and 12-h-aged PLEDs (squares: as-prepared PLEDs; circles: 12-h-aged PLEDs).

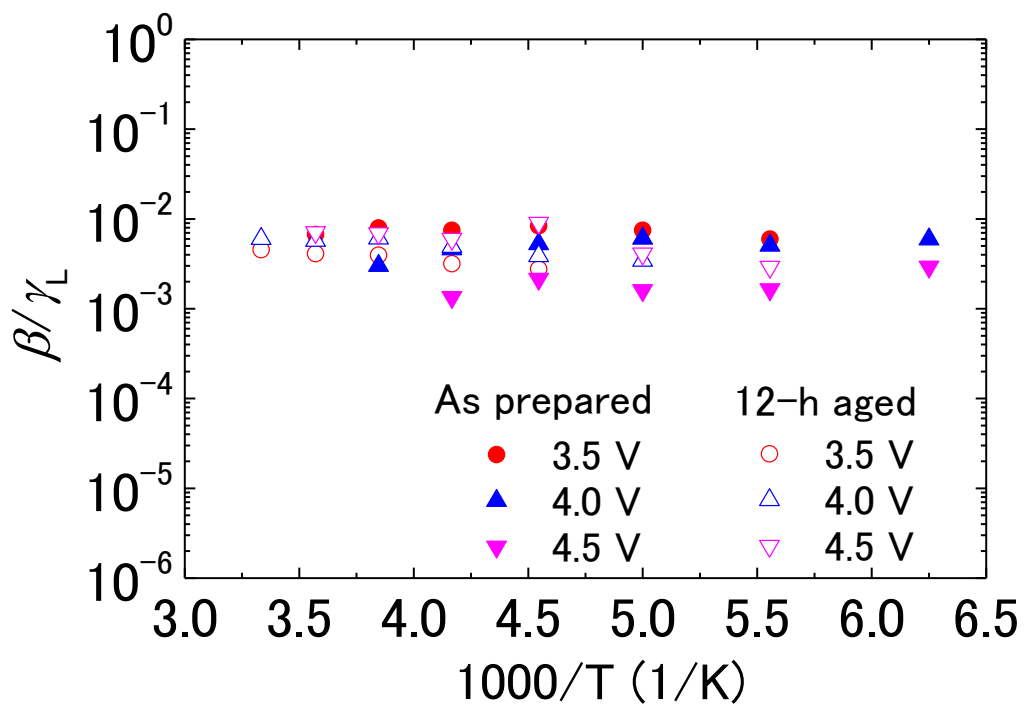

**FIG 10.** Plots of the Langevin recombination reduction factor $\beta/\gamma_L$ versus temperature under various applied voltages in the as-prepared and 12-h-aged PLEDs (closed symbols: as-prepared PLEDs; open symbols: 12-h-aged PLEDs).

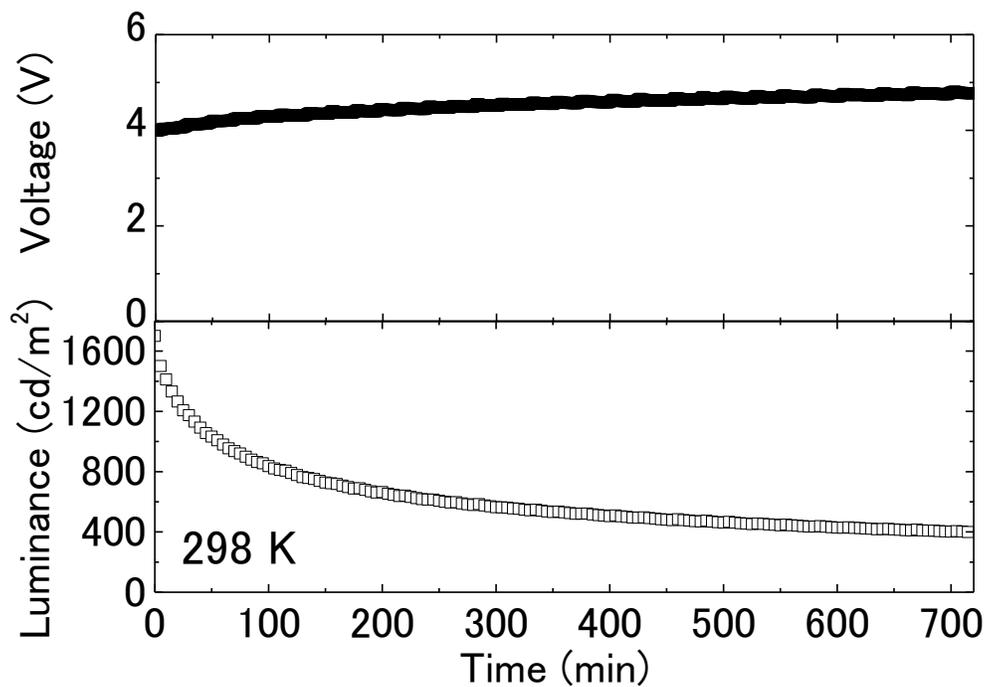

**FIG 11.** Temporal variations in the driving voltage and luminance of the PLEDs when operated at 20 mA cm$^{-2}$.

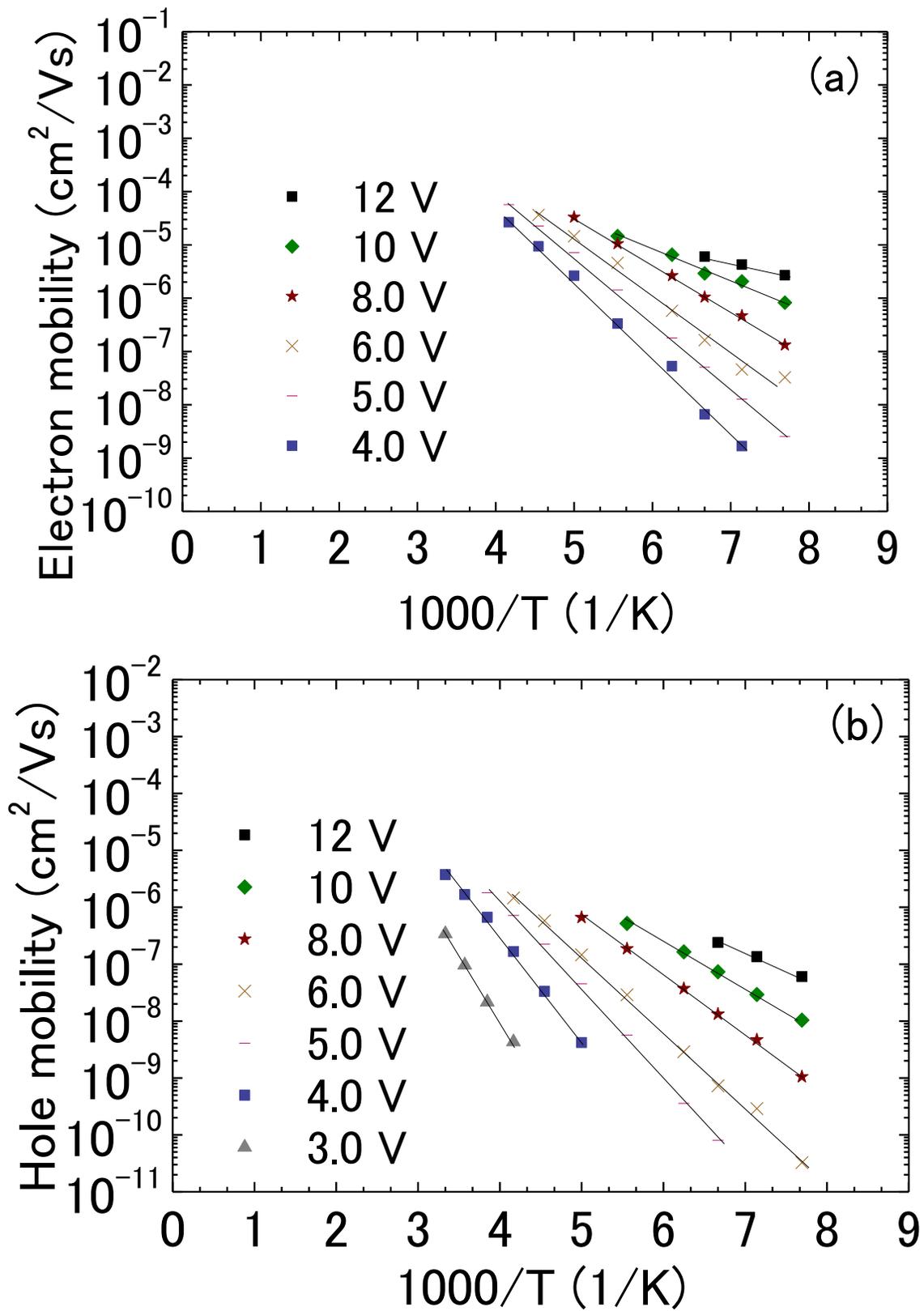

**FIG 12.** Plots of (a) electron drift mobility ($\mu_n$) and (b) hole drift mobility ($\mu_p$) versus temperature under various applied voltages in the 12-h-aged F8BT-based PLEDs.

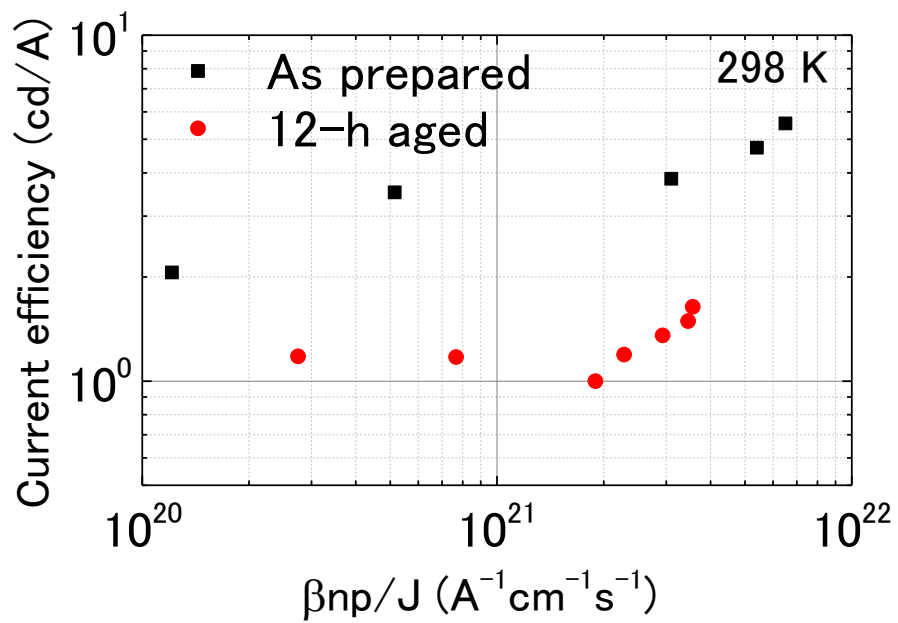

**FIG 13.** Plots of current efficiency versus the ratio of the bimolecular recombination rate to the current density ($\beta np/J$) in the as-prepared and 12-h-aged PLEDs (squares: as-prepared PLEDs; circles: 12-h-aged PLEDs).

# Supplementary Material

# Full characterization of electronic transport properties in working polymer light-emitting diodes via impedance spectroscopy

Makoto Takada,[1] Takashi Nagase,[1,2] Takashi Kobayashi,[1,2] and Hiroyoshi Naito[1,2,2)]

[1]*Department of Physics and Electronics, Osaka Prefecture University, Sakai 599-8531, Japan*
[2]*The Research Institute for Molecular Electronic Devices, Osaka Prefecture University, Sakai 599-8531, Japan*

## Calculation of localized state energy distributions

We show numerically here that the localized tail state energy distributions from the mobility edges can be determined from the electric field dependences of the activation energy of the drift mobility measured using impedance spectroscopy. The expression for the complex impedance $Z_1$ of a single-injection space-charge-limited diode in the presence of localized states has been derived on the basis of trap-controlled band transport and is given by the following infinite series [S1-S3]:

$$Z_1 = 6\psi R_i \sum_{k=0}^{\infty} \frac{1}{k+3} \frac{\Gamma(\psi+1)}{\Gamma(\psi+k+2)} \left(\frac{\psi}{\delta}\right)^k (-j\Omega)^k \tag{16}$$

where $\Omega \, (=\omega t_{t0})$ is the transit angle, $\omega \, (=2\pi f)$ is the angular frequency of the small ac voltage, and $t_{t0} \, [= 4d^2/(3\mu_m V_{dc})]$ is the transit time of the charge carriers in the absence of the localized states; $d$ is the thickness of the semiconductor layer, $\mu_m$ is the microscopic mobility of the charge carriers, $V_{dc}$ is the applied dc bias voltage, and $R_i$ is the differential resistance of the diodes, which is expressed as

$$R_i = \frac{4}{9} \frac{d^3}{\varepsilon_r \varepsilon_0 \mu_m \delta V_{dc} S}, \tag{17}$$

where $\varepsilon_r$ is the dielectric constant of the semiconductor, $\varepsilon_0$ is the vacuum permittivity, and $S$ is the active area. $\delta$ and $\psi$ are trapping parameters that are given as follows:

$$\delta = \left[1 + \int_{E_v}^{E_c} \frac{\gamma_c(E)}{\gamma_t(E)} dE\right]^{-1} \tag{18}$$

$$\psi(\omega) = \left[1 + \int_{E_v}^{E_c} \frac{\gamma_c(E)}{\gamma_t(E) + j\omega} dE\right] \delta \tag{19}$$

where $E_C$ is the conduction band mobility edge, $E_V$ is the valence band mobility edge, $\gamma_t(E) \{= \nu \exp[-(E_C - E)/(kT)]\}$ is the release rate from the localized state located at energy $E$, and $\gamma_c(E)dE \, [\approx c_n N_t(E)dE]$ is the capture rate of the localized state at energy $E$; $\nu \ (= c_n N_C)$ is the attempt-to-escape frequency, $c_n$ is the electron capture coefficient, $N_t(E)$ is the energetic distribution of the localized tail state density, and $N_C$ is the effective density of states in the conduction band.

In the following, we calculated the complex impedance in the electron transport case, but the same results are obtained in the hole transport case. We assume localized tail state distributions above the $E_C$ value of Eq. (S5):

$$N_t(E) = N_0 \exp\left[-\left(\frac{E_C - E}{E_0}\right)^A\right], \tag{20}$$

where $N_0$ is the density of localized states at $E_C$, $E_0$ is the energy scale of the localized states, and $A$ is the exponent of the distribution. At $A=1$, Eq. (S5) gives an exponential distribution, while at $A=2$, Eq. (S5) gives a Gaussian distribution.

The following physical quantities were used in the numerical calculations: $\varepsilon_r = 3$, $\varepsilon_0 = 8.85 \times 10^{-12}$ F m$^{-1}$, $c_n = 1.0 \times 10^{-8}$ cm$^3$ s$^{-1}$, $\mu_m = 10^{-1}$ cm$^2$V$^{-1}$s$^{-1}$, $N_C = 2.5 \times 10^{19}$ cm$^{-3}$, $N_0 = 1.0 \times 10^{21}$ cm$^{-3}$ eV$^{-1}$, $d = 100$ nm, and $S = 4$ mm$^2$. These physical quantities are reasonable as values for organic semiconductors [S4]. The frequency dependences of the capacitance calculated using Eq. (S1) in the case where $A = 1$ and $E_0 = 30$ meV are shown in Fig. S1. Dispersive transport was observed in the case where the values of $E_0$ in Eq. (S5) are used in the present calculation.

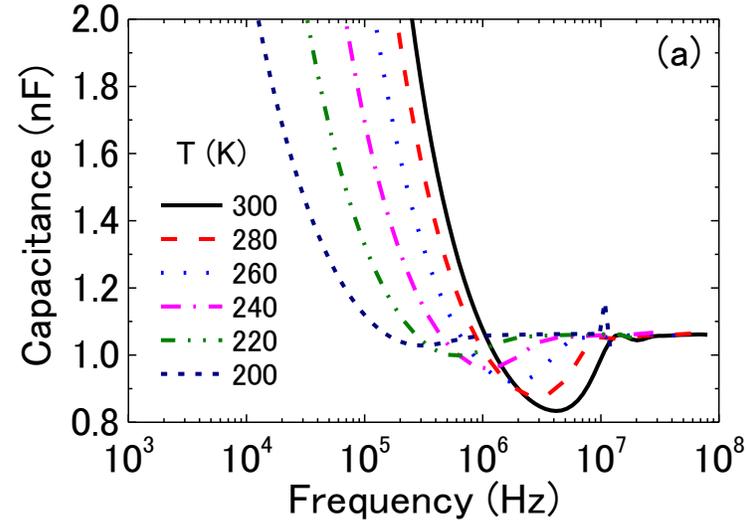

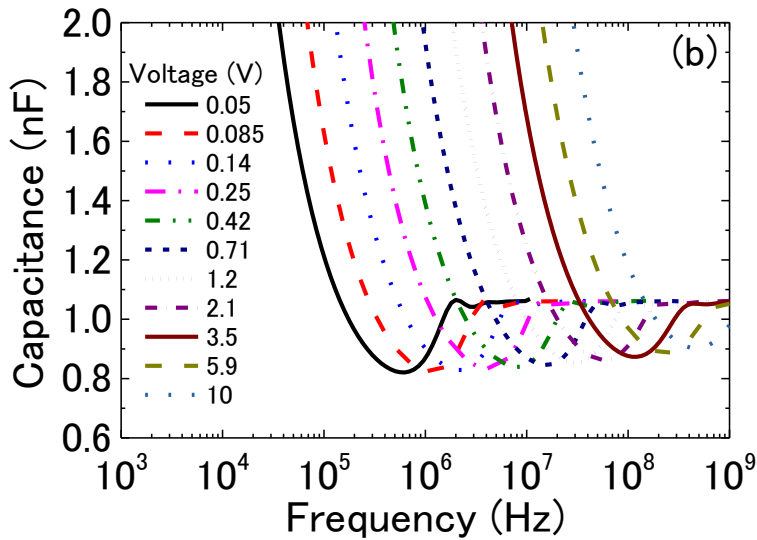

Fig. S1. Frequency dependences of capacitance when calculated using Eq. (S1) for (a) various temperatures at $V_{dc}$ = 0.25 V and (b) various applied voltages at 300 K. The following physical quantities were used in the calculations: $\varepsilon_r$ = 3, $\varepsilon_0$ = 8.85×10⁻¹² F m⁻¹, $c_n$ = 1.0×10⁻⁸ cm³ s⁻¹, $\mu_m$ = 10⁻¹ cm²V⁻¹s⁻¹, $N_C$ = 2.5×10¹⁹ cm⁻³, $N_0$ = 1.0×10²¹ cm⁻³ eV⁻¹, $d$ = 100 nm, $S$ = 4 mm², $A$ = 1, and $E_0$ = 30 meV.

The temperature dependences of the drift mobilities determined from Fig. S1 using the $-\Delta B$ method are shown in Fig. S2. The activation energies for the drift mobility at different values of $V_{dc}$ are obtained from Fig. S2. The electric field dependences of the activation energy ($E_a$) values of the drift mobilities are shown in Fig. S3. The electric field dependences of $E_a$ are dependent on the values of $E_0$ in Eq. (S5) (note that the electric field dependences of $E_a$ are observed in the dispersive transport case [39, S6]).

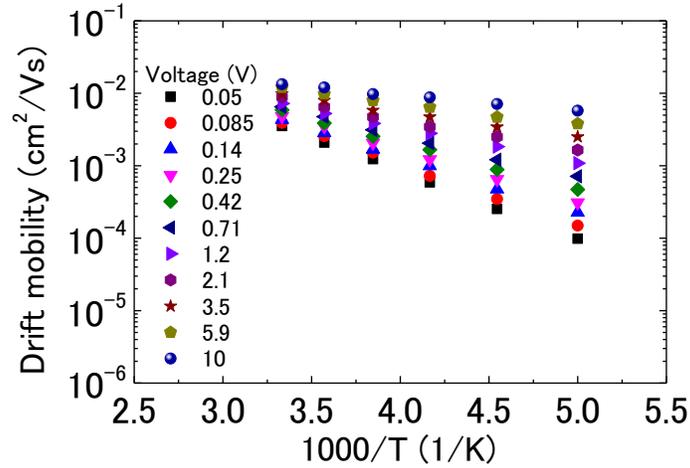

Fig. S2. Temperature dependences of the drift mobilities at various applied voltages. The following physical quantities were used in the calculations: $\varepsilon_r$ = 3, $\varepsilon_0$ =8.85×10⁻¹² F m⁻¹, $c_n$ = 1.0×10⁻⁸ cm³ s⁻¹, $\mu_m$ = 10⁻¹ cm²V⁻¹s⁻¹, $N_C$ = 2.5×10¹⁹ cm⁻³, $N_0$ = 1.0×10²¹ cm⁻³ eV⁻¹, $d$ = 100 nm, $S$ = 4 mm², $A$ = 1, and $E_0$ = 30 meV.

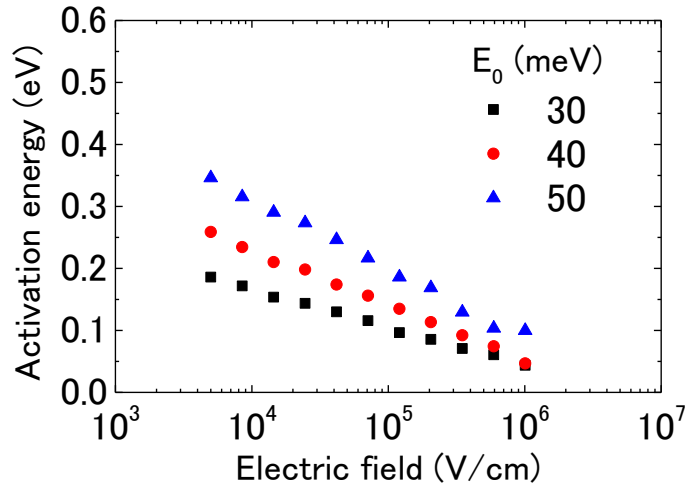

Fig. S3. Plots of the activation energies of the drift mobilities versus the electric field calculated for $E_0$ values of 30, 40, and 50 meV.

The localized tail state distributions from the conduction band mobility edges determined using Eq. (9) from the electric field dependences of the values of $E_a$ in Fig. S3 [S7] are shown in Fig. S4. These calculated distributions show good agreement with the input distributions.

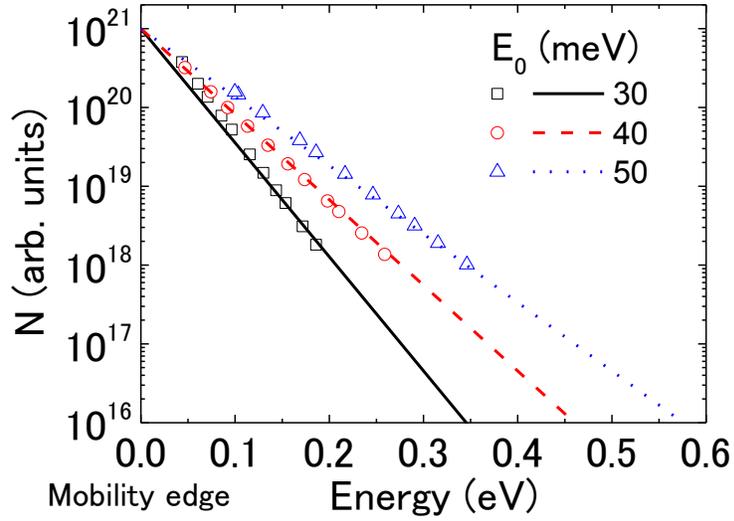

Fig. S4. Localized tail state distributions from the conduction band mobility edges, as determined from the electric field dependences of the values of $E_a$ in Fig. S3 for various values of $E_0$. The localized tail state distributions (open symbols) show good agreement with the input localized tail state distributions (solid, dashed, and dotted lines).

We also calculated the localized tail state distributions in the cases where $A = 1.5$ and 2 in Eq. (S5). The localized tail state distributions from the conduction band mobility edges for various values of $A$ are shown in Fig. S5. These calculated distributions show good agreement with the input distributions in both cases. These results demonstrate that the method used in this work can be applied to analysis of the electric field dependence of $E_a$ when determined using impedance spectroscopy in the cases of representative tail state distributions.

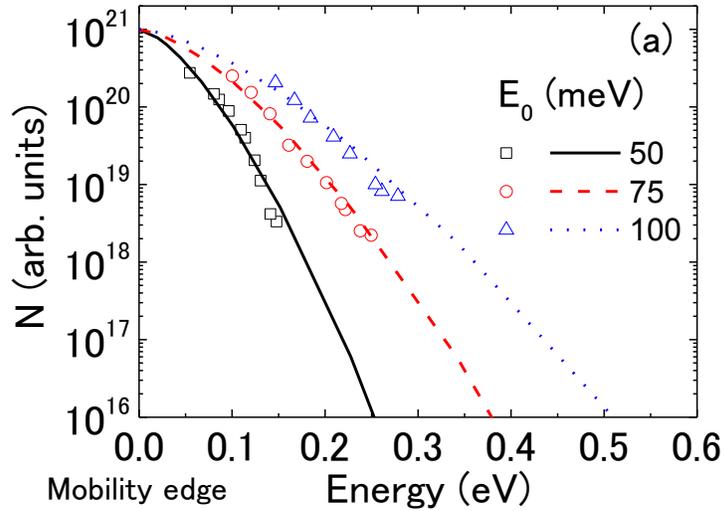

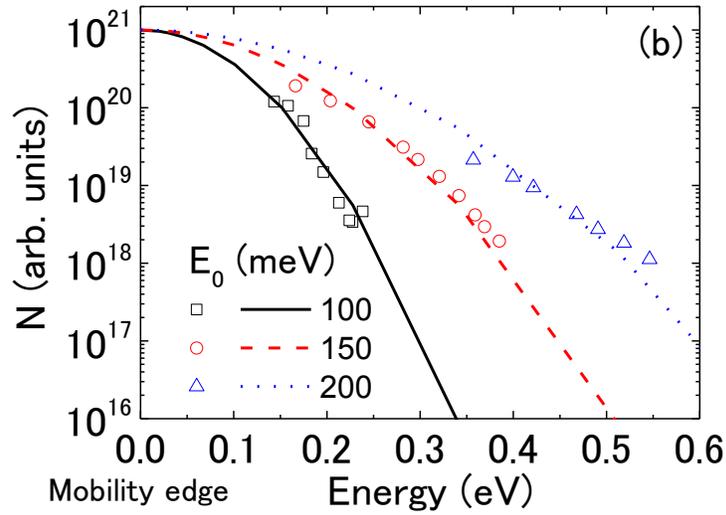

Fig. S5. Localized tail state distributions from the conduction band mobility edges for various values of $E_0$ in the cases where (a) $A = 1.5$ and (b) $A = 2$ (open symbols: calculated distributions; solid, dashed, and dotted lines: input distributions).